# *Estimating Chicago's tree cover and canopy height using multi-spectral satellite imagery*

**John Francis**
The Alan Turing Institute &
University College London

**Stephen Law**
University College London &
The Alan Turing Institute

## Abstract

Information on urban tree canopies is fundamental to mitigating climate change [1] as well as improving quality of life [2]. Urban tree planting initiatives face a lack of up-to-date data about the horizontal and vertical dimensions of the tree canopy in cities. We present a pipeline that utilizes LiDAR data as ground-truth and then trains a multi-task machine learning model to generate reliable estimates of tree cover and canopy height in urban areas using multi-source multi-spectral satellite imagery for the case study of Chicago.

## 1 Introduction

Major American cities such as New York, Los Angeles, Boston, and Chicago have set forth tree planting initiatives as part of larger efforts to mitigate climate change, improve quality of life, and promote environmental equity. One of the primary issues policymakers face when deciding how to allocate tree planting resources is a lack of high quality, up-to-date datasets about the urban canopy. This paper proposes a novel pipeline for generating estimates of two urban canopy measures, tree cover and canopy height, in Chicago for timepoints when high-quality data is unavailable. Previous research has utilized machine learning (ML) approaches to predict tree canopy [3] but have yet to leverage these techniques to create detailed estimates of urban tree cover and canopy height as demonstrated in this project.

## 2 Data on the Urban Canopy

Three main techniques have been previously developed to generate estimates of urban canopies. Surveying techniques have been used, for example, by Morton Arboretum in Chicago to produce a tree census. This tree census counted the number of trees in 268 selected plots, and then extrapolated those numbers to the entire city, estimating that Chicago contains about 3,997,000 trees [4]. As it would be impossible for humans to physically count the number of trees in an area as large as Chicago, the exact number of trees remains unknown and can only be estimated.

An alternative to surveying techniques is airborne Light Detection and Ranging (LiDAR). LiDAR utilizes light beams to create a cloud of millions of points that can be accurate within a few centimeters [5]. Numerous algorithms exist for detecting and measuring trees from LiDAR point clouds that can be used to generate three-dimensional representations of the urban canopy. Despite this, LiDAR can only capture a point in time, and the collection of LiDAR data, especially over large areas, is quite expensive, so researchers are often relegated to using outdated data.

To get around a lack of consistent data collection, ML techniques can be used to generate accurate, up-to-date estimates of urban canopies. Weinstein et al. [6] used a convolutional neural network to identify individual trees through image segmentation in a California forest using RGB (red, green, and blue) images. In addition to RGB images, multi-spectral (MS) imagery has been shown to aid in predictive models. Wang Li et al. [3] successfully used a ML model to predict forest canopy height from MS imagery in a mountainous region of China. Furthermore, previous research has tried to predict relative building and vegetation height and semantic segmentation masks simultaneously [7,8]. Despite these efforts, studies have yet to estimate tree cover and canopy height in an urban setting using MS satellite images.

This paper focuses on multiple measures of the urban canopy, tree cover and canopy height, because alone, neither horizontal nor vertical measures of the urban canopy capture the whole effect trees have on the environment. In general, larger trees tend to have a greater environmental effect than smaller trees, with numerous smaller trees often unable to match the effects of a single large tree [9, 10]. By accounting for both a horizontal and vertical measure, a more accurate assessment of the urban canopy's impact on climate indicators can eventually be quantified.

## 3 Estimating Tree Cover and Canopy Height

Chicago LiDAR point cloud data from 2017 was retrieved from the Illinois Geospatial Data Clearinghouse [11]. Additional MS satellite imagery was used from the National Agriculture Imagery Program (NAIP) and the Sentinel-2 satellite program. Four-band NAIP RBG and NIR data at 1-m resolution was collected from the US Geological Survey's earth explorer for 2017 and 2019. Sentinel-2 data from 2017 and 2019 was collected from the Sentinel Hub's Earth observation browser, with four bands at 10-m resolution, and six bands at 20-m resolution. A methodology similar to Roussel et al. [12] was used to extract ground truth tree cover and canopy height measures from the LiDAR data. Details on how the input data was prepared can be found in the appendix.

### 3.2 Training the UNet Multi-Task Model

This paper followed a multi-task (MT) learning approach, utilizing the UNet architecture which has achieved good performance on various pixel-level tasks [13,14,15,16]. Three tasks, predicting if a pixel is part of a tree (tree mask), estimating how tall a pixel represents (pixel height), and a third auxiliary task predicting whether a pixel represents impervious space (NDVI<0) were included together to enable better generalizations on individual tasks. The UNet architecture consists of an encoding path and a decoding path that share representations to increase the output's resolution [17].

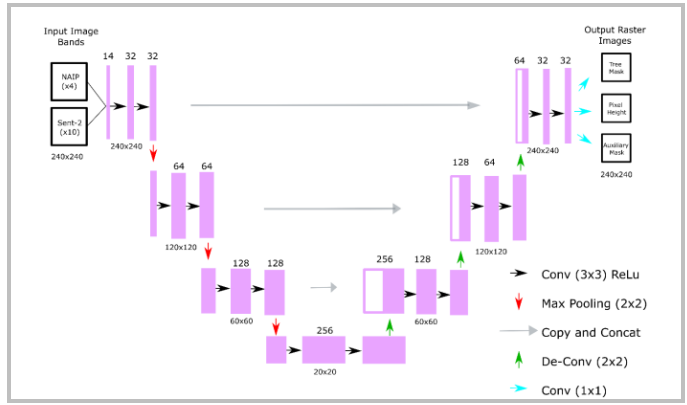

Figure 1. UNet model architecture.

The primary MT model used in this paper can be seen in Figure 1. The output of the decoding path is fed into three separate convolutional layers, one with a linear activation (pixel height) and two with sigmoid activation (tree mask, auxiliary mask). To retrieve canopy height, the predicted tree mask was applied to the pixel height layer to only retain the height of pixels determined by the model to be trees. An Adam optimizer algorithm was used to train models with mean squared error loss used for pixel height and with Jaccard distance loss used for the two binary masks to account for class imbalance in the data [18]. In total, 9,535 240x240 images from 2017 were used to train the model, with 25% of these images held out as a testing set. The input images for the model consisted of the 14 MS bands from the NAIP and Sentinel 2 satellite images. In addition to the primary model shown in Figure 1, separate models for comparison were run with just the RGB bands of the NAIP images, single-task versions of the model for each output, and versions where only the encoding features of the model were shared by the three tasks, with separate decoding layers for each output. The model was trained and evaluated solely using data from 2017. It was then used to predict tree cover and canopy height for 2019 where ground truth data does not exist. To evaluate model performance, Intersection over Union (IoU) is used for the tree mask and the impervious surface mask, while Mean Absolute Error (MAE) is used for pixel height.

## 4 Results

Ten UNet models were run to determine which method was best able to locate trees and determine their height. Table 1 shows the results of these models. The model that was best able to locate trees (IoU=.647) was the MT model with fully shared layers. This was unexpected as the MT model with partially shared layers (encoding only) contained nearly twice as many parameters and allowed for more features specific to the individual tasks. Because the tasks are all closely related, the features most relevant to each individual task may simply be the shared features. Pixel height was best predicted by the UNet model looking at pixel height alone within about five percent of the observed value. It is possible that pixel height predictions suffered in the MT models because the weighting scheme was more focused on generating an accurate prediction of tree location. Notably, while the MT models using MS image bands achieved better results for all three tasks, nearly comparable results were found when using only the RGB bands of the NAIP data.

Table 1: UNet model results

| Model | Bands Used | Tree Mask IoU | Height MAE | Auxiliary IoU |
| --- | --- | --- | --- | --- |
| Tree Mask Alone | RGB Only | .475 | - | - |
| Tree Mask Alone | 14 MS Bands | .476 | - | - |
| Pixel Height Alone | RGB Only | - | .063 | - |
| **Pixel Height Alone** | **14 MS Bands** | - | **.050** | - |
| Auxiliary Mask | RGB Only | - | - | .131 |
| Auxiliary Mask | 14 MS Bands | - | - | .747 |
| MT Fully Shared | RGB Only | .614 | .099 | .878 |
| **MT Fully Shared** | **14 MS Bands** | **.647** | .085 | **.940** |
| MT Partially Shared | RGB Only | .621 | .070 | .884 |
| MT Partially Shared | 14 MS Bands | .642 | .072 | .934 |

While the IoU of .647 and the MAE of .050 indicated relatively good predictions of tree cover and canopy height, it was important to visually inspect the results to ensure that the metrics provided an accurate assessment of model performance. Figure 2 shows the ground truth raster layers generated by the analysis of the LiDAR data, the RGB bands of the NAIP satellite image that fed into the model, and the predicted output raster layers generated by the models for tree cover and canopy height among the 2017 test data. Tree cover seems to mimic the ground truth data well, although there may be a slight overinflation of tree size in some spots. Additionally, it appears that some smaller trees may have been missed, while some small shrubbery may have been misidentified as trees. This is to be expected as even a human looking through the satellite images would have a difficult time capturing all the pixels that contain trees. For pixel height, the model clearly struggled with some of the taller buildings; however, it is important to remember that only the height of pixels determined to be trees are interpreted in final canopy height estimates. Among locations that are clearly trees from the satellite data, the model seemed to appropriately predict height values.

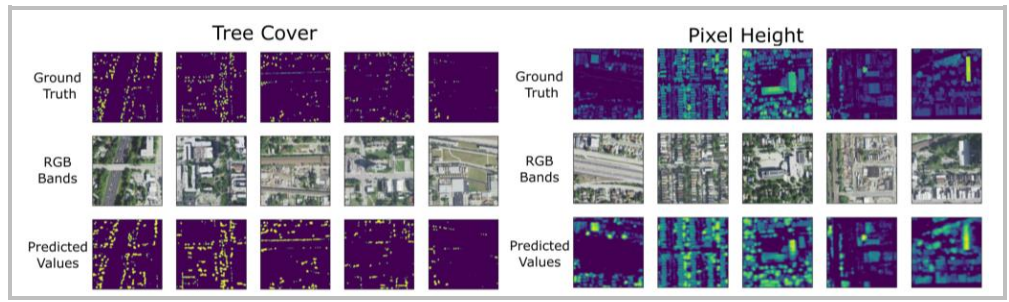

Figure 2: 2017 tree cover and pixel height predictions.

**4.1 Predicting 2019 Tree Cover and Canopy Height**

Utilizing the model trained in 2017, we inferred the 2019 tree cover and canopy height metrics for Chicago and estimated a total city-wide cover of 5.9%. This is a slight increase from the ground truth 2017 LiDAR data which calculated the city-wide cover to be about 4.8%. Notably, these estimates are much lower than results from Chicago's 2020 tree census which estimated the canopy cover to be nearly 16% when including shrubs [4]. This paper's estimates may be more reliable than the larger tree cover estimates proposed via survey techniques as they are based on the highly accurate 2017 LiDAR data. The ten sample images shown in Figure 2 all provide examples of areas with less than 10% of pixels identified as trees. Figure 3 shows the 2019 UNet predictions at 46,149 census blocks. The maps here indicate that the areas of highest tree cover are concentrated primarily in the northern part of the city, as well as within the many parks that line Chicago's eastern coast.

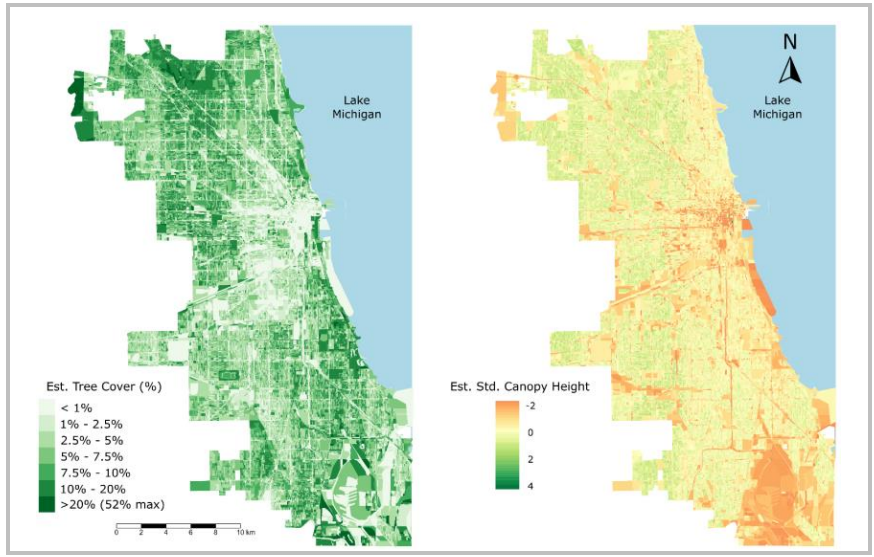

Figure 3: Estimated tree cover and canopy height.

## 5   Conclusions and Future Work

This paper provides a novel pipeline for estimating tree cover and canopy height. Utilizing ML, specifically the UNet architecture with MS imagery, researchers and policy makers are presented a method with which to generate up-to-date and accurate measures of the urban canopy. With better data, better decisions can be made about where to plant new trees in cities to maximize climate benefits while improving equity and quality of life for as many people as possible. The results from this study provide confidence that ML methodologies can generate usable estimates of the urban canopy moving forward. Additionally, many studies utilizing satellite imagery across different domains for various ML tasks only use RGB image bands, often because RGB bands are the only bands available at high resolutions. This study provides initial evidence that including MS bands in ML models, even when these bands are at lower resolutions, can lead to slight increases in predictive capacity. Urban tree canopies are constantly evolving, so to keep up with these ever-changing environments, policymakers need to arm themselves with the highest quality data. When and where LiDAR data collection is unavailable, ML methods provide a promising alternative for researchers and governments to generate estimates of the urban canopy.

Moving forward, newer ML models could be easily integrated into this pipeline, while more resources could allow for the usage of higher resolution MS imagery to further improve predictions. Additionally, there is a need to test model generalizability geographically leveraging global LiDAR datasets (e.g. GEDI) as well as including additional auxiliary tasks (e.g. species type, above ground carbon estimates) which can provide useful information for urban tree planting policies and initiatives. By leveraging ML techniques to generate high quality data, public officials will be given more confidence that their decisions will have strong and lasting positive impacts on communities.

## Acknowledgements and Funding

This work was supported by Towards Turing 2.0 under the EPSRC Grant EP/W037211/1 and The Alan Turing Institute. This work was completed as part of the Social and Geographic Data Science MSc program within the University College London's Department of Geography. Special thanks to Mat Disney for his advice on tree metrics and the use of LiDAR data.

convolutional networks with jaccard distance. IEEE transactions on medical imaging, 36(9), 1876-1886.

[19] Zhao, Z., Wang, H., Wang, C., Wang, S., & Li, Y. (2019). Fusing LiDAR data and aerial imagery for building detection using a vegetation-mask-based connected filter. IEEE Geoscience and Remote Sensing Letters, 16(8), 1299-1303.

[20] Dalponte, M. & Coomes, D.A. (2016). Tree-centric mapping of forest carbon density from airborne laser scanning and hyperspectral data. Methods Ecol Evol 7:1236–1245.

## Appendix

**Preparing the UNet input data**

The LiDAR point cloud data consisted of 1,131 2500x2500 foot tiles with a derived nominal pulse spacing of one point every 0.35 meters. First, a vegetation mask was created using the NIR and Red bands of the NAIP images, creating a normalized difference vegetation index (NDVI) raster layer as calculated in Zhao et al. [19]. The point cloud was then masked, keeping only points that were vertically aligned with NDVI values above .05. Masking the point cloud not only speeds up calculations, but also prevents buildings and non-biological objects from being classified as trees. Points below six feet and above 80 feet were filtered out to ignore small shrubbery and any incidental non-vegetation points (e.g., birds). A canopy height model was then generated using a pitfree algorithm which allows for individual tree detection using a local maximum filter. Next, trees were segmented based on the Dalponte and Coomes algorithm [20]. From this process, two raster layers were generated, one with binary values if a pixel was identified as being part of a tree, while the other raster layer contained the average max height of each pixel. These raster layers were then mosaiced together and stacked on top of the Sentinel-2 and NAIP data which were all projected to the extent and resolution of the NAIP 1-m data. These raster stacks were then cut into 240x240 pixel patches to be used as the input for a convolutional neural network.